\documentclass[aps,prl,twocolumn,superscriptaddress]{revtex4-2}
\usepackage{amsmath,amssymb}
\usepackage{booktabs} 
\usepackage{graphicx}
\usepackage{xcolor}
\usepackage{array}
\usepackage{hyperref}
\usepackage{tikz}
\usetikzlibrary{positioning, fit, shapes}
\usepackage{pgfplots}

\begin{document}
\setcounter{secnumdepth}{1}

\title{How graviton mass shapes angular correlations in a relic gravitational-wave background}

\author{Hongguang Zhang}
\email{zhanghg@bjtu.edu.cn}
\affiliation{Weihai Research Institute, Beijing Jiaotong University, Weihai 264401, China}

\author{Zhengyang Qiu}
\email{zhengyangqiu@bjtu.edu.cn}
\affiliation{Weihai Research Institute, Beijing Jiaotong University, Weihai 264401, China}

\author{Sirui Wang}
\email{siruiwang@bjtu.edu.cn}
\affiliation{Weihai Research Institute, Beijing Jiaotong University, Weihai 264401, China}

\author{Jiayi Wei}
\email{weijiayi@bjtu.edu.cn}
\affiliation{Weihai Research Institute, Beijing Jiaotong University, Weihai 264401, China}

\author{Jing Zhang}
\email{zhangjing@bjtu.edu.cn}
\affiliation{Weihai Research Institute, Beijing Jiaotong University, Weihai 264401, China}

\begin{abstract}
A nonzero graviton mass makes waves emitted at the same epoch travel different distances at different momenta. A finite band therefore samples a shell that is both closer to the observer and thicker. We derive the distribution of source radii and use it to calculate the band angular-power ratio $C_\ell^{BB}(m_{\rm g})/C_\ell^{BB}(0)$ for a relic stochastic gravitational-wave background. For a common source field with a scale-invariant spatial spectrum, a thin shell has the same angular power at every radius. Radial averaging cannot exceed this power, so any suppression relative to the massless single-epoch spectrum is due to shell broadening. We also ask whether source mixing can mimic massive propagation across two frequency bands. For two independent emission episodes whose mixing fractions are fixed across multipoles, massless propagation requires the low-band auto-power and the cross-power, each normalized by the high-band auto-power, to lie on a straight line when plotted against one another. This holds for arbitrary angular spectra of the two sources. A massive benchmark near the low-band propagation threshold violates this relation. A joint massless fit with power-law sources nevertheless gives an rms logarithmic residual of $0.013$ over $2\leq\ell\leq10$.
\end{abstract}

\maketitle

\section{Introduction}

Gravitational waves produced in the early universe can form a relic stochastic gravitational-wave background (SGWB)~\cite{Maggiore2000,Caprini2018}. Its angular power spectrum describes how the intensity varies across the sky on different angular scales. Contaldi developed a line-of-sight description of SGWB anisotropies~\cite{Contaldi2017}. Cusin, Pitrou, and Uzan calculated angular correlations of the energy density of an astrophysical background~\cite{Cusin2018}, and Bartolo et al.\ studied anisotropies of a cosmological background~\cite{Bartolo2019}. These studies show how the angular pattern observed today depends on both the distribution of the sources and the propagation of the waves.

A nonzero graviton mass makes the propagation speed depend on momentum~\cite{Hinterbichler2012,deRham2014,deRhamEtAl2017}. Will showed how this dispersion changes the waveform of an inspiralling binary~\cite{Will1998}, an effect subsequently tested with compact-binary observations~\cite{LIGOScientific2016,GWTC3Tests}. For stochastic backgrounds, previous studies have examined how a graviton mass changes pulsar timing correlations~\cite{Lee2010} and the frequency spectrum of cosmological gravitational waves~\cite{Gumrukcuoglu2012}. A related effect arises for relic neutrinos. Bisnovatyi-Kogan and Seidov studied how neutrino mass affects propagation distances in an expanding universe~\cite{BisnovatyiKoganSeidov1984}. Dodelson and Vesterinen calculated how the distribution of neutrino momenta gives the neutrino last-scattering surface a finite radial thickness~\cite{DodelsonVesterinen2009}.

The neutrino result motivates a question about the gravitational-wave sky. Consider waves emitted at the same cosmic time in a homogeneous expanding universe. If the graviton is massless, waves reaching us today have all traveled the same comoving distance. Their sources lie on a thin shell around the observer, which we call the effective last-scattering surface whether or not an actual scattering event occurred. If the graviton has mass, lower-momentum waves travel more slowly and reach us from smaller radii. A finite band then samples a shell that is both displaced inward and broadened. Along each line of sight, the measured intensity averages source fluctuations over the thickness of this shell. How does this averaging change the angular power, and under what conditions can its effect be separated from that of shifting the shell?

To answer this question, we calculate the distance traveled by each mode and weight it by its contribution to the present-day energy density in the selected band. The resulting distribution of source radii defines a radial window. We project a common field of fractional source-intensity fluctuations through this window to obtain $C_\ell^{BB}(m_{\rm g})/C_\ell^{BB}(0)$ for each graviton mass. A scale-invariant spatial source spectrum allows the two geometric effects to be separated: the angular power of a thin shell is independent of its radius, and averaging over radii cannot raise the power above that thin-shell value.

The same propagation effect raises a second question. If two frequency bands sample different angular patterns, how much of that difference can be explained by the sources themselves? For massless propagation from a single epoch, bands tracing the same fractional source field sample the same shell and have identical maps. We first allow the source amplitude and tilt to differ between bands and test the resulting predictions for their auto- and cross-spectra. We then allow two independent emission episodes to contribute in different proportions to the two bands. Even massless waves can give imperfectly correlated maps in this case. We derive a relation that the three massless spectra must satisfy together at different multipoles and test it against massive propagation in two fixed observed bands.

Section~\ref{sec:window} derives the propagation distance and the radial window. Section~\ref{sec:width} relates the shell width to the spreads in momentum and emission time. Section~\ref{sec:angular} derives the angular power spectrum, the thin-shell upper bound, and the massless relation between frequency bands. Section~\ref{sec:example} illustrates the single-band suppression and the corresponding graviton mass scales. Section~\ref{sec:cross_band} compares the auto- and cross-spectra of two fixed frequency bands and tests the massless two-source consistency relation. Section~\ref{sec:conclusion} summarizes the results.

\section{Massive Propagation and the Radial Window}
\label{sec:window}

We assume free propagation in a spatially flat, unperturbed FLRW universe, set $a_0=1$, and use natural units $c=\hbar=1$. The graviton mass $m_{\rm g}$ is constant, with dispersion relation $E^2=p^2+m_{\rm g}^2$. We use this as a propagation model for tensor radiation~\cite{deRhamEtAl2017}; the source weights are specified separately. A mode with present-day physical momentum $p_0$ has momentum $p(a)=p_0/a$ and, by differentiating $E$ with respect to $p$, group velocity $v_{\rm g}=p/\sqrt{p^2+m_{\rm g}^2}$. In geometric optics, its comoving displacement satisfies $\mathrm{d}\chi=v_{\rm g}\mathrm{d}t/a$. Using $\mathrm{d}t=\mathrm{d}a/[aH(a)]$ gives the comoving propagation distance, as in the corresponding calculation for relic neutrinos~\cite{DodelsonVesterinen2009}:
\begin{equation}
\chi(p_0,a_{\rm em};m_{\rm g})
=
\int_{a_{\rm em}}^1
\frac{\mathrm{d}a}{a^2H(a)}
\frac{p_0}{\sqrt{p_0^2+a^2m_{\rm g}^2}}.
\label{eq:generalized_vg}
\end{equation}
For a massive mode, lower momentum means slower propagation and a smaller source radius. Later emission also gives a smaller radius because less time is available for travel. Thus, both momentum and emission time determine which part of the source distribution is sampled today.

We next assign a weight to each contribution. Let $\mathcal E(p_0,a_{\rm em})\,\mathrm{d}\ln p_0\,\mathrm{d}a_{\rm em}$ be its present-day energy density divided by the present critical density $\rho_{c,0}$. This quantity includes the production history, redshifting, and dilution. Integrating over emission time gives the energy density parameter per logarithmic momentum interval:
\begin{equation}
\Omega_{{\rm gw},0}(p_0)
=
\int_0^1
\mathrm{d}a_{\rm em}\,
\mathcal E(p_0,a_{\rm em}).
\label{eq:extended_omega_gw}
\end{equation}
A nonnegative weight $R_B(p_0)$ selects the band $B$. The normalized distribution of momenta and emission epochs contributing to this band is
\begin{equation}
P_{p,B}(p_0,a_{\rm em})
=
\frac{
R_B(p_0)\mathcal E(p_0,a_{\rm em})
}{
\mathcal N_B
}.
\label{eq:generalized_momentum_distribution}
\end{equation}
Here $\mathcal N_B\equiv\int\mathrm{d}\ln p_0\int_0^1\mathrm{d}a_{\rm em}\,R_B(p_0)\mathcal E(p_0,a_{\rm em})$, with $0<\mathcal N_B<\infty$. The distribution therefore integrates to unity with measure $\mathrm{d}\ln p_0\,\mathrm{d}a_{\rm em}$. Its weight is set by the energy received today.

If the band is specified in observed frequency, we use $R_B(p_0;m_{\rm g})=R_B^{(f)}[f(p_0;m_{\rm g})]$, where $f(p_0;m_{\rm g})=\sqrt{p_0^2+m_{\rm g}^2}/(2\pi)$. A fixed frequency band then selects different momenta for different masses. This dependence must be included when comparing mass hypotheses for the same observed band; we leave it implicit in $P_{p,B}$.

Equation~\eqref{eq:generalized_vg} assigns each pair $(p_0,a_{\rm em})$
a propagation distance $\chi_{\rm prop}\equiv\chi(p_0,a_{\rm em};m_{\rm g})$.
Grouping the weights from Eq.~\eqref{eq:generalized_momentum_distribution}
by this distance gives the radial window per logarithmic distance interval,
\begin{equation}
\widetilde W_B(\chi;m_{\rm g})
=\int\! \mathrm{d}\ln p_0 \int_0^1\! \mathrm{d}a_{\rm em}\,
P_{p,B}\,\delta\!\left(\ln\frac{\chi}{\chi_{\rm prop}}\right).
\label{eq:generalized_visibility_def}
\end{equation}
Here $P_{p,B}$ denotes $P_{p,B}(p_0,a_{\rm em})$.
The delta function places each contribution at its propagation distance.
If the probability at $\chi=0$ is zero, the window is normalized:
$\int_0^\infty \mathrm{d}\ln\chi\,\widetilde W_B(\chi;m_{\rm g})=1$.

We measure its width by $\sigma_{\ln\chi,B}^2\equiv\langle(\ln\chi-\langle\ln\chi\rangle_B)^2\rangle_B$, where $\langle\cdot\rangle_B$ denotes an average over $P_{p,B}$. Logarithms of distances are taken relative to a fixed reference length, and we assume finite logarithmic moments. For a narrow shell, $\sigma_{\ln\chi,B}$ is approximately its thickness divided by its radius.

\section{Radial Broadening from Momentum and Emission Time}
\label{sec:width}

Setting $m_{\rm g}=0$ in Eq.~\eqref{eq:generalized_vg} gives the
massless propagation distance
$\chi_0(a_{\rm em})=\int_{a_{\rm em}}^1 \mathrm{d}a/[a^2H(a)]$,
where $H(a)=H_0[\Omega_r a^{-4}+\Omega_m a^{-3}+\Omega_\Lambda]^{1/2}$.
When $m_{\rm g}\ll p_0$ for the momenta carrying appreciable weight,
the group velocity is $v_{\rm g}\simeq 1-a^2m_{\rm g}^2/(2p_0^2)$.
Substituting this expansion into Eq.~\eqref{eq:generalized_vg} gives
\begin{equation}
\chi \simeq \chi_0-\frac{m_{\rm g}^2}{2p_0^2}I_1.
\label{eq:generalized_chi_expansion}
\end{equation}
Here $I_1=\int_{a_{\rm em}}^1 \mathrm{d}a/H(a)$, evaluated at the
same emission epoch as $\chi_0$. The omitted terms are of order
$(m_{\rm g}/p_0)^4$ relative to $\chi_0$. Defining
$\mathcal A_{\rm prop}=I_1/\chi_0$, the fractional shift is
$(\chi-\chi_0)/\chi_0\simeq-\mathcal A_{\rm prop}m_{\rm g}^2/(2p_0^2)$.
At a fixed emission epoch, lower momenta therefore correspond to smaller
shell radii, so a band containing a range of momenta samples a shell
of finite thickness.

To calculate the width, define the representative values $p_{{\rm eff},B}=\exp[\langle\ln p_0\rangle_B]$ and $a_{\rm eff}=\exp[\langle\ln a_{\rm em}\rangle_B]$, with momentum expressed in a fixed reference unit. For a distribution concentrated around these values, expanding $\ln\chi$ from Eq.~\eqref{eq:generalized_vg} to first order in both variables gives
\begin{equation}
\delta\ln\chi
\simeq
A_B\,\delta\ln p_0
+
B_B\,\delta\ln a_{\rm em}.
\label{eq:distance_linearization}
\end{equation}
Here $\delta$ denotes a deviation from the corresponding mean, $A_B\equiv(\partial\ln\chi/\partial\ln p_0)_{\rm eff}$, and $B_B\equiv(\partial\ln\chi/\partial\ln a_{\rm em})_{\rm eff}$. The coefficients describe how strongly distance responds to each input. This approximation requires them to vary little across the contributing distribution, but it does not require a small mass.

Writing $v(a)=p_0/\sqrt{p_0^2+a^2m_{\rm g}^2}$, we have $\partial v/\partial\ln p_0=v(1-v^2)$. Differentiating Eq.~\eqref{eq:generalized_vg} with respect to $\ln p_0$ and dividing by $\chi$ gives
\begin{equation}
\frac{\partial\ln\chi}{\partial\ln p_0}
=
\frac{1}{\chi}
\int_{a_{\rm em}}^1
\frac{\mathrm{d}a}{a^2H(a)}
v(a)\bigl[1-v^2(a)\bigr].
\label{eq:exact_momentum_sensitivity}
\end{equation}
This response is nonnegative: increasing momentum increases the distance traveled. It vanishes in the massless limit and is small whenever the mode remains highly relativistic. Differentiating Eq.~\eqref{eq:generalized_chi_expansion} with respect to $\ln p_0$ and dividing by $\chi_0$ gives the leading term,
\begin{equation}
\frac{\partial\ln\chi}{\partial\ln p_0}
\simeq
\mathcal A_{\rm prop}(a_{\rm em})
\left(\frac{m_{\rm g}}{p_0}\right)^2.
\label{eq:small_mass_sensitivity}
\end{equation}
The response to emission time follows from differentiating the lower limit of Eq.~\eqref{eq:generalized_vg} and multiplying by $a_{\rm em}/\chi$,
\begin{equation}
\frac{\partial\ln\chi}{\partial\ln a_{\rm em}}
=
-\frac{v(a_{\rm em})}
{\chi\,a_{\rm em}H(a_{\rm em})}.
\label{eq:emission_time_sensitivity}
\end{equation}
Its sign is negative because later emission shortens the travel interval. For massless propagation, the response is $-1/[\chi_0(a_{\rm em})a_{\rm em}H(a_{\rm em})]$. A spread in emission time can therefore broaden the shell even when the graviton mass is zero.

Squaring Eq.~\eqref{eq:distance_linearization} and averaging gives the radial variance. If logarithmic momentum and emission scale factor are uncorrelated, the result is
\begin{equation}
\sigma_{\ln\chi,B}^2
\simeq
A_B^2\sigma_{\ln p,B}^2
+
B_B^2\sigma_{\ln a,B}^2.
\label{eq:generalized_sigma_coupling}
\end{equation}
Here $\sigma_{\ln p,B}^2$ and $\sigma_{\ln a,B}^2$ are their variances under $P_{p,B}$. The first term measures the effect of different speeds; the second measures the effect of different travel times. Correlations add $2A_BB_B\operatorname{Cov}_B(\ln p_0,\ln a_{\rm em})$ to the right-hand side. Since $A_B\geq0$ and $B_B\leq0$, a positive correlation can reduce the width: higher momentum can partly compensate for later emission.

For a single emission epoch, Eq.~\eqref{eq:generalized_sigma_coupling} has $\sigma_{\ln a,B}=0$, so $\sigma_{\ln\chi,B}\simeq A_B\sigma_{\ln p,B}$. Substituting Eq.~\eqref{eq:small_mass_sensitivity} for $A_B$ gives
\begin{equation}
\sigma_{\ln\chi,B}
\simeq
\mathcal A_{\rm prop}(a_{\rm eff})
\left(\frac{m_{\rm g}}{p_{{\rm eff},B}}\right)^2
\sigma_{\ln p,B}.
\label{eq:small_mass_radial_width}
\end{equation}
Thus, in the relativistic and narrow-band limit, the width grows as the squared mass-to-momentum ratio multiplied by the momentum width. The factor $\mathcal A_{\rm prop}$ accounts for cosmic expansion. As the momentum width tends to zero, the shell becomes thin even though its radius remains smaller than the massless radius.

There is also a bound that does not require either approximation. Equation~\eqref{eq:exact_momentum_sensitivity} is a positive weighted average of $1-v^2$, so $0\leq\partial\ln\chi/\partial\ln p_0\leq1$. At a fixed emission epoch, a difference in logarithmic distance can therefore never exceed the corresponding difference in logarithmic momentum. Let $p_0$ and $p_0'$ be independent draws from the same band-weighted momentum distribution, with corresponding propagation distances $\chi$ and $\chi'$. Squaring the preceding bound and averaging over the two draws gives $\langle[\ln(\chi/\chi')]^2\rangle_B\leq\langle[\ln(p_0/p_0')]^2\rangle_B$. Since the mean squared difference between two independent draws equals twice their variance, the two sides are $2\sigma_{\ln\chi,B}^2$ and $2\sigma_{\ln p,B}^2$, respectively. Taking square roots gives $\sigma_{\ln\chi,B}\leq\sigma_{\ln p,B}$ whenever $\ln p_0$ has finite variance.

\section{From the Radial Window to the Angular Power Spectrum}
\label{sec:angular}

We assume that all contributing momentum channels and emission epochs trace the same statistically homogeneous and isotropic fractional source-intensity field $\delta S(\boldsymbol x)$. They therefore share a spatial pattern of intensity fluctuations. This assumption does not require phase coherence between gravitational-wave frequencies. Averaging $\delta S$ along each line of sight with the window in Eq.~\eqref{eq:generalized_visibility_def} gives
\begin{equation}
\delta_B(\hat{\boldsymbol n})
=
\int \mathrm{d}\ln\chi\,
\widetilde W_B(\chi;m_{\rm g})
\delta S(\chi\hat{\boldsymbol n}).
\label{eq:projected_source_field}
\end{equation}
A thin window samples the field at one radius along each direction. A thick window averages contributions from several radii.

The projection uses the same spherical-Bessel geometry as line-of-sight calculations of CMB and SGWB anisotropies~\cite{Seljak1996,Contaldi2017}. In Eq.~\eqref{eq:projected_source_field}, we expand each Fourier mode in spherical harmonics. Averaging the squared harmonic coefficients over source realizations gives
\begin{equation}
C_\ell^{BB}(m_{\rm g})
=
4\pi\int_0^\infty
\frac{\mathrm{d}k}{k}\,
\mathcal P_{\delta S}(k)
\left|\Delta_\ell^B(k;m_{\rm g})\right|^2,
\label{eq:generalized_Cl_window_projection}
\end{equation}
where
\begin{equation}
\Delta_\ell^B(k;m_{\rm g})
=
\int \mathrm{d}\ln\chi\,
\widetilde W_B(\chi;m_{\rm g})
j_\ell(k\chi).
\label{eq:generalized_window_transfer}
\end{equation}
The superscript $BB$ denotes the autocorrelation of band $B$. The dimensionless spectrum $\mathcal P_{\delta S}(k)$ describes spatial fluctuations of the source intensity. Its wavenumber $k$ sets the size of those fluctuations and differs from the gravitational-wave momentum $p_0$, which determines the propagation speed. The function $j_\ell(k\chi)$ converts a spatial Fourier mode at radius $\chi$ into a contribution to angular multipole $\ell$.

Substituting Eq.~\eqref{eq:generalized_visibility_def} into Eq.~\eqref{eq:generalized_window_transfer} and integrating over the delta function gives
\begin{equation}
\begin{aligned}
\Delta_\ell^B(k;m_{\rm g})
&=
\int \mathrm{d}\ln p_0
\int_0^1 \mathrm{d}a_{\rm em}\,
P_{p,B}(p_0,a_{\rm em};m_{\rm g})
\\
&\quad\times
j_\ell\!\left[k\chi(p_0,a_{\rm em};m_{\rm g})\right].
\end{aligned}
\label{eq:generalized_exact_momentum_transfer}
\end{equation}
Here we display the possible mass dependence of the normalized weight explicitly. Even for a fixed observed frequency selection, the corresponding momentum weights can change with mass.

Equation~\eqref{eq:generalized_exact_momentum_transfer} connects the propagation calculation to the angular spectrum. For each mass, the distance $\chi$ determines where each momentum and emission contribution samples the source field, while $P_{p,B}$ determines its weight. Because the channels trace a common source field, their transfer functions are combined before taking the squared magnitude in Eq.~\eqref{eq:generalized_Cl_window_projection}. This retains their statistical correlations and the full shape of the radial window.

Dividing Eq.~\eqref{eq:generalized_Cl_window_projection} by its massless value, when that value is nonzero, gives
\begin{equation}
\frac{C_\ell^{BB}(m_{\rm g})}{C_\ell^{BB}(0)}
=
\frac{
\displaystyle\int_0^\infty \frac{\mathrm{d}k}{k}\,
\mathcal P_{\delta S}(k)
\left|\Delta_\ell^B(k;m_{\rm g})\right|^2
}{
\displaystyle\int_0^\infty \frac{\mathrm{d}k}{k}\,
\mathcal P_{\delta S}(k)
\left|\Delta_\ell^B(k;0)\right|^2
}.
\label{eq:general_angular_power_ratio}
\end{equation}
The numerator and denominator use the same assumed source-fluctuation spectrum. The band selection and the prescription for the source weights must also be specified consistently. This ratio gives the change in angular power associated with a mass hypothesis under those choices.

Two effects contribute to this change. Moving the shell changes the spatial scales sampled at a given angle. Giving the shell a finite thickness averages fluctuations over different radii, whose angular contributions can partly cancel. The source spectrum and the full radial window determine their combined effect. Consequently, the width alone does not determine the angular power, and a general source spectrum need not give suppression relative to massless propagation.

To isolate the effect of thickness, we choose
$\mathcal P_{\delta S}(k)=A_S$, with $A_S>0$.
This is a scale-invariant spatial spectrum of source-intensity fluctuations; it does not specify the frequency dependence of the background energy density. For $\ell\geq1$ and $\chi>0$, the Weber--Schafheitlin integral, together with
$j_\ell(z)=\sqrt{\pi/(2z)}J_{\ell+1/2}(z)$, gives~\cite{DLMF}
\begin{equation}
\int_0^\infty
\frac{\mathrm{d}k}{k}\,
j_\ell^2(k\chi)
=
\frac{1}{2\ell(\ell+1)}.
\label{eq:thin_shell_identity}
\end{equation}
For a thin shell, Eq.~\eqref{eq:generalized_window_transfer} reduces to $j_\ell(k\chi)$. Using Eq.~\eqref{eq:thin_shell_identity} in Eq.~\eqref{eq:generalized_Cl_window_projection} then gives
$C_{\ell,\mathrm{thin}}=2\pi A_S/[\ell(\ell+1)]$,
independent of its radius. In this source model, moving the shell does not change the angular power, allowing us to isolate the effect of its thickness.

For a normalized, nonnegative radial window, applying the Cauchy--Schwarz inequality to Eq.~\eqref{eq:generalized_window_transfer} gives
\begin{equation}
\left|\Delta_\ell^B(k;m_{\rm g})\right|^2
\leq
\int \mathrm{d}\ln\chi\,
\widetilde W_B(\chi;m_{\rm g})
j_\ell^2(k\chi).
\label{eq:projection_averaging_bound}
\end{equation}
Multiplying Eq.~\eqref{eq:projection_averaging_bound} by $4\pi A_S$, integrating over $\mathrm{d}k/k$, and using Eq.~\eqref{eq:thin_shell_identity} yields
$C_\ell^{BB}(m_{\rm g})\leq C_{\ell,\mathrm{thin}}$.
For a single emission epoch, the massless window is a thin shell, so
\begin{equation}
0\leq
\frac{C_\ell^{BB}(m_{\rm g})}{C_\ell^{BB}(0)}
\leq1.
\label{eq:angular_power_ratio_bound}
\end{equation}
Any reduction in this case comes from radial averaging rather than shell displacement. With extended emission, the massless window can already be thick, so the thin-shell bound alone does not determine which of the massive and massless spectra is larger.

For emission at a single epoch, setting $m_{\rm g}=0$ in Eq.~\eqref{eq:generalized_vg} gives the same radius $\chi_0$ at every frequency. If two bands trace the same fractional source field, their normalized windows therefore give the same sky map by Eq.~\eqref{eq:projected_source_field}. Their angular powers are equal for any common spatial source spectrum.

\section{Single-Band Angular Suppression and Graviton Mass Scales}
\label{sec:example}

For the numerical illustration, we use $(\Omega_r,\Omega_m,\Omega_\Lambda)=(10^{-4},0.3,0.6999)$, a single emission epoch $a_{\rm eff}=10^{-10}$, and the common source field with $\mathcal P_{\delta S}(k)=A_S$. We hold the shape of the band-weighted momentum distribution fixed when comparing masses. A comparison in a fixed observed frequency band would instead require recalculating the momentum weights for each mass.

The early emission epoch makes the single-epoch approximation useful. At $a_{\rm eff}=10^{-10}$, the magnitude of the massless emission-time response is $1/[\chi_0a_{\rm eff}H(a_{\rm eff})]\simeq3.09\times10^{-9}$. A moderate logarithmic spread around this epoch changes only a very small initial part of the travel distance. For production concentrated there, with negligible late emission, this contribution is much smaller than the dispersion widths considered below. We therefore use a single epoch to isolate the momentum-dependent propagation effect.

The momentum distribution is Gaussian in $x=\ln(p_0/p_{{\rm eff},B})$. Writing $s_B=\sigma_{\ln p,B}$, its normalized measure is
\begin{equation}
\mathrm{d}\mu_B
=
\frac{e^{-x^2/(2s_B^2)}}{\sqrt{2\pi}s_B}
\,\mathrm{d}x,
\qquad
-\infty<x<\infty.
\label{eq:gaussian_momentum_measure}
\end{equation}
The representative momentum $p_{{\rm eff},B}$ is its geometric mean. With $q=m_{\rm g}/p_{{\rm eff},B}$, substituting $p_0=p_{{\rm eff},B}e^x$ into Eq.~\eqref{eq:generalized_vg} gives
\begin{equation}
\chi_q(x)
\equiv
\int_{a_{\rm eff}}^1
\frac{\mathrm{d}a}{a^2H(a)}
\frac{1}{\sqrt{1+a^2q^2e^{-2x}}}.
\label{eq:numerical_exact_distance}
\end{equation}
We use this full relation for both the radial width and the angular spectrum, including $q\gtrsim1$. With geometric mean radius $\bar\chi_B=\exp[\int\mathrm{d}\mu_B\ln\chi_q(x)]$, using the fixed reference length for logarithms, averaging the squared logarithmic distance deviation from Eq.~\eqref{eq:numerical_exact_distance} with the weights in Eq.~\eqref{eq:gaussian_momentum_measure} gives
\begin{equation}
\sigma_{\ln\chi,B}^2
=
\int\mathrm{d}\mu_B
\left[
\ln\frac{\chi_q(x)}{\bar\chi_B}
\right]^2.
\label{eq:numerical_radial_variance}
\end{equation}

\begin{figure}[t]
\centering
\includegraphics[width=\linewidth]{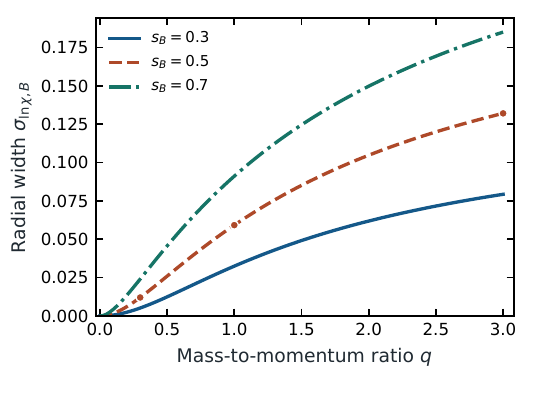}
\caption{Effective radial broadening $\sigma_{\ln\chi,B}$ as a function of the mass-to-momentum ratio $q$ for logarithmic momentum widths $\sigma_{\ln p,B}=0.3,0.5,0.7$. The curves use the full propagation relation and Gaussian logarithmic momentum distributions, with a fixed emission epoch $a_{\rm eff}=10^{-10}$. Markers on the $\sigma_{\ln p,B}=0.5$ curve identify the benchmarks $q=0.3,1,3$ used in Fig.~\ref{fig:damping_plot}.}
\label{fig:broadening_plot}
\end{figure}

Figure~\ref{fig:broadening_plot} shows the width over $0\leq q\leq3$ for $s_B=0.3,0.5,0.7$. The width vanishes at $q=0$, when all momenta travel the same distance. Over the plotted range, increasing $q$ or the momentum width gives a thicker shell. In the small-mass, narrow-band limit, Eq.~\eqref{eq:small_mass_radial_width} gives $\sigma_{\ln\chi,B}\simeq\mathcal A_{\rm prop}q^2s_B$, with $\mathcal A_{\rm prop}\simeq0.1644$. For $s_B=0.5$, the full calculation gives widths of approximately $0.0121$, $0.0592$, and $0.1321$ at $q=0.3,1,3$, respectively. We use these three cases for the angular calculation.

\begin{figure}[t]
\centering
\includegraphics[width=\linewidth]{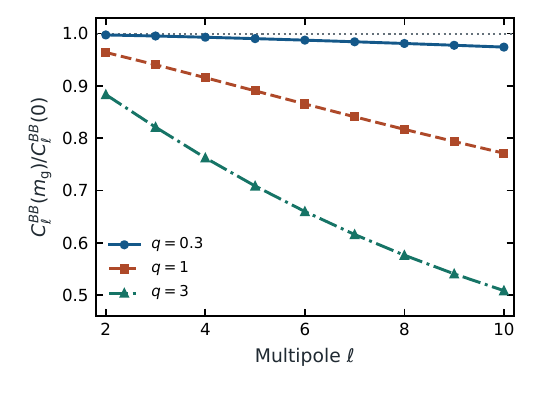}
\caption{Angular-spectrum ratios $C_\ell^{BB}(m_{\rm g})/C_\ell^{BB}(0)$ for the scale-invariant source spectrum $\mathcal P_{\delta S}(k)=A_S$, logarithmic momentum width $\sigma_{\ln p,B}=0.5$, and mass-to-momentum ratios $q=0.3,1,3$. The calculation uses full propagation from $a_{\rm eff}=10^{-10}$ and the complete angular projection. Markers indicate integer multipoles, and the dotted line denotes the massless reference.}
\label{fig:damping_plot}
\end{figure}

Using the Gaussian weights in Eq.~\eqref{eq:generalized_exact_momentum_transfer} gives $\Delta_\ell^B(k;m_{\rm g})=\int\mathrm{d}\mu_B\,j_\ell[k\chi_q(x)]$. Evaluating the massless denominator of Eq.~\eqref{eq:general_angular_power_ratio} with Eq.~\eqref{eq:thin_shell_identity} gives
\begin{equation}
\frac{C_\ell^{BB}(m_{\rm g})}{C_\ell^{BB}(0)}
=
2\ell(\ell+1)
\int_0^\infty
\frac{\mathrm{d}k}{k}\,
\left|\Delta_\ell^B(k;m_{\rm g})\right|^2.
\label{eq:numerical_angular_ratio}
\end{equation}
The source amplitude $A_S$ cancels. The idealized scale-invariant spectrum extends over $0<k<\infty$, and the projected integrals converge for $\ell\geq1$. The common distance factor $H_0^{-1}$ also cancels from this ratio and from the logarithmic radial variance. These quantities depend on the relative spread of radii, rather than their overall distance scale.

Figure~\ref{fig:damping_plot} shows the ratios over $2\leq\ell\leq10$ for $s_B=0.5$. At $\ell=10$, they are approximately $0.974$, $0.771$, and $0.508$ for $q=0.3,1,3$. The corresponding power reductions are about $3\%$, $23\%$, and $49\%$. Larger mass ratios give wider windows and stronger suppression over the plotted range. Suppression also increases toward higher multipoles in this range, where finer angular structure is more affected by averaging contributions from different radii. These values describe a reduction in the angular contrast of the fractional intensity map, not a loss of total background energy density.

To assign physical masses to these examples, define the representative frequency by $2\pi f_B=\sqrt{p_{{\rm eff},B}^2+m_{\rm g}^2}$ in natural units. Substituting $p_{{\rm eff},B}=m_{\rm g}/q$, solving for $m_{\rm g}$, and restoring $\hbar$ gives
\begin{equation}
m_{\rm g}
=
2\pi\hbar f_B
\frac{q}{\sqrt{1+q^2}}.
\label{eq:mass_scale_estimate}
\end{equation}
At fixed $q$, cosmology, emission epoch, and momentum-distribution shape, multiplying the frequency and mass by the same factor leaves the speed history unchanged. It therefore leaves the radial window and angular-power ratio unchanged within the propagation model. This gives curves proportional to $m_{\rm g}\propto f_B$ for a fixed angular response.

Motivated by pulsar timing observations of a nanohertz background~\cite{NANOGrav2023,CPTA2023}, we choose $f_B=3\,\mathrm{nHz}$ as a representative frequency. At this frequency, the cases $q=0.3,1,3$ correspond to masses of approximately $3.57\times10^{-24}$, $8.77\times10^{-24}$, and $1.18\times10^{-23}\,\mathrm{eV}$. Table~\ref{tab:frequency_mass_benchmarks} gives the $q=1$ conversion at representative PTA, LISA~\cite{LISA2017}, and Advanced LIGO (aLIGO)~\cite{AdvancedLIGO2015} frequencies. These frequencies provide reference energy scales for the conversion.

\begin{table}[t]
\centering
\caption{Representative frequencies and graviton masses for $q=1$. These are kinematic benchmarks, not observational bounds.}
\label{tab:frequency_mass_benchmarks}
\small
\renewcommand{\arraystretch}{1.2}
\setlength{\tabcolsep}{5pt}
\begin{tabular}{lcc}
\hline\hline
Detector & $f_B$ & $m_{\rm g}$ [eV] \\
\hline
PTA   & $3\,\mathrm{nHz}$ & $8.77\times10^{-24}$ \\
LISA  & $3\,\mathrm{mHz}$ & $8.77\times10^{-18}$ \\
aLIGO & $100\,\mathrm{Hz}$ & $2.92\times10^{-13}$ \\
\hline\hline
\end{tabular}
\end{table}

\begin{figure}[t]
\centering
\includegraphics[width=\linewidth]{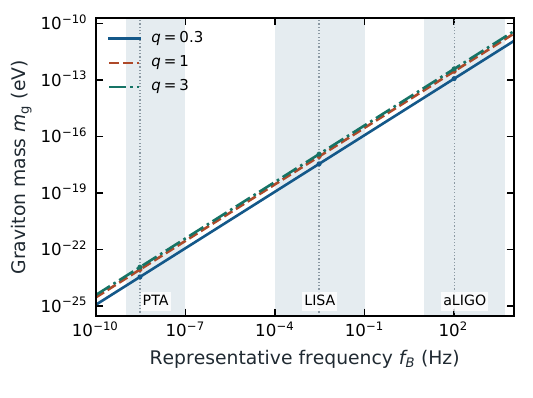}
\caption{Frequency--mass correspondence for fixed mass-to-momentum ratios $q=0.3,1,3$. The curves follow Eq.~\eqref{eq:mass_scale_estimate} and retain the source model and momentum width of Fig.~\ref{fig:damping_plot}, associated with angular-power ratios $C_{10}^{BB}(m_{\rm g})/C_{10}^{BB}(0)\simeq0.974,0.771,0.508$, respectively. Shading indicates approximate detector bands; vertical lines and markers identify the representative frequencies. These are equal-response curves within the adopted propagation model, not sensitivity curves or mass limits. At the lowest frequencies, the curves are formal kinematic extensions; their equal-response interpretation requires the geometric-optics condition to hold over the momentum range carrying appreciable weight.}
\label{fig:frequency_mass_plot}
\end{figure}

Figure~\ref{fig:frequency_mass_plot} extends the conversion over $10^{-10}\leq f_B/\mathrm{Hz}\leq10^4$. Shading marks approximate bands of $10^{-9}$--$10^{-7}\,\mathrm{Hz}$ for PTA, $10^{-4}$--$10^{-1}\,\mathrm{Hz}$ for LISA, and $10$--$5\times10^3\,\mathrm{Hz}$ for aLIGO. Each curve compares different constant-mass hypotheses with the same $q$ and angular response. At fixed mass, increasing frequency instead reduces $q$ and approaches the massless limit.

The interpretation of these curves requires geometric optics. The representative modes are relativistic at emission, so we require $p_{{\rm eff},B}/[a_{\rm eff}H(a_{\rm eff})]\gg1$. For $H_0=70\,\mathrm{km\,s^{-1}\,Mpc^{-1}}$ and $f_B=3\,\mathrm{nHz}$, this ratio is approximately $80$, $59$, and $26$ for $q=0.3,1,3$. It becomes of order unity near the lowest plotted frequencies. There the curves provide only the algebraic frequency--mass conversion; following propagation from the fixed emission epoch with a group velocity is no longer reliable. For a finite band, the geometric-optics condition must hold over the momenta carrying appreciable weight.

The masses displayed here are model benchmarks, not observational bounds. A mass inference must keep the measured frequency band fixed and include the source spectrum, production history, signal amplitude, and measurement uncertainties. For a PTA, the mass-dependent timing response~\cite{Lee2010} must also be combined with the sky-multipole response~\cite{Mingarelli2013} and anisotropy uncertainties~\cite{Depta2025}. Independent mass constraints also matter~\cite{deRhamEtAl2017,GWTC3Tests}. For example, the GWTC-4.0 dispersion analysis gives $m_{\rm g}\leq1.92\times10^{-23}\,\mathrm{eV}$ at $90\%$ credibility~\cite{GWTC4Parameterized}. The PTA examples lie below this particular bound, whereas the $q=1$ masses at the LISA and aLIGO reference frequencies exceed it for the same constant-mass dispersion model. Those high-frequency entries serve only to illustrate the conversion in Eq.~\eqref{eq:mass_scale_estimate}.

\section{Angular Power and Correlations across Frequency Bands}
\label{sec:cross_band}

We first adopt the cosmology and single emission epoch of Sec.~\ref{sec:example} and fix the observed bands to $L=[2,4]\,\mathrm{nHz}$ and $H=[5,7]\,\mathrm{nHz}$. In each band we take the received energy weight to be uniform per logarithmic frequency interval, $\mathrm{d}\mu_B=\mathrm{d}\ln f/\ln(f_{\max,B}/f_{\min,B})$. With $c=1$ and $h=2\pi\hbar$, the dispersion relation gives $p_0(f;m_{\rm g})=\sqrt{(hf)^2-m_{\rm g}^2}$. Thus a fixed weight in $\ln f$ has a mass-dependent density in $\ln p_0$, since $\mathrm{d}\ln f/\mathrm{d}\ln p_0=p_0^2/(p_0^2+m_{\rm g}^2)$. All masses in Table~\ref{tab:two_band} satisfy $m_{\rm g}<hf_{\min,L}$.

Inserting this momentum into Eq.~\eqref{eq:generalized_vg} gives the radius $\chi_f\equiv\chi(p_0(f;m_{\rm g}),a_{\rm eff};m_{\rm g})$ at each observed frequency. Averaging the Bessel function in Eq.~\eqref{eq:generalized_exact_momentum_transfer} over the selected band then gives
\begin{equation}
\Delta_\ell^B(k;m_{\rm g})=\int_B\mathrm{d}\mu_B(f)\,j_\ell(k\chi_f).
\label{eq:fixed_freq_transfer}
\end{equation}
We use Eq.~\eqref{eq:generalized_Cl_window_projection} for each band with $\mathcal P_{\delta S}(k)=A(k/k_*)^n$, where $k_*$ is a fixed pivot. At each graviton mass, we compare the two bands through the angular-power ratio $C_\ell^{LL}(m_{\rm g})/C_\ell^{HH}(m_{\rm g})$, assuming nonzero high-band power. The common source amplitude $A$ cancels in this ratio. For massless propagation, the equal spectra derived in Sec.~\ref{sec:angular} give $C_\ell^{LL}(0)/C_\ell^{HH}(0)=1$. For massive propagation, the two bands sample different radial windows. If the source spectrum is tilted, differences in both mean radius and radial spread can affect the ratio. To evaluate the power, we expand the squared band average as a double frequency integral and perform the inner integral over the product of two Bessel functions using Eqs.~10.22.56 and 10.22.57 of Ref.~\cite{DLMF}. We vary $n$ independently for each mass to test the sensitivity of the predictions to the common tilt.

\begin{table}[t]
\centering
\small
\setlength{\tabcolsep}{7pt}
\renewcommand{\arraystretch}{1.3}
\begin{tabular}{@{}cccc@{}}
\toprule
\addlinespace[3pt]
\shortstack{$m_{\rm g}[\mathrm{eV}]$}
& $C_2^{LL}/C_2^{HH}$
& $C_5^{LL}/C_5^{HH}$
& $C_{10}^{LL}/C_{10}^{HH}$ \\
\addlinespace[4pt]
\midrule
0.0                  & 1.000 & 1.000 & 1.000 \\
$3.6\times 10^{-24}$ & 1.000 & 0.998 & 0.995 \\
$6.0\times 10^{-24}$ & 0.995 & 0.983 & 0.954 \\
$8.0\times 10^{-24}$ & 0.959 & 0.880 & 0.760 \\
\bottomrule
\end{tabular}
\caption{Illustrative low-to-high band angular-power ratios for a
scale-invariant source spectrum ($n=0$). Values are rounded to three
decimal places.}
\label{tab:two_band}
\end{table}

Table~\ref{tab:two_band} shows the scale-invariant case. At $\ell=10$, the low-band power is about $5\%$ below the high-band power for $6.0\times10^{-24}\,\mathrm{eV}$ and about $24\%$ below it for $8.0\times10^{-24}\,\mathrm{eV}$. Their predicted ratios remain separated when the common tilt varies over $[-0.2,0.2]$. The largest effect occurs near the propagation threshold: the lower band starts at $hf\simeq8.27\times10^{-24}\,\mathrm{eV}$, close to the largest mass. The propagation distance changes rapidly with frequency near this edge, so the lower band samples a wider range of radii. At $2\,\mathrm{nHz}$ the heaviest benchmark has $p_0\simeq2.1\times10^{-24}\,\mathrm{eV}$, giving $p_0/[a_{\rm eff}\hbar H(a_{\rm eff})]\simeq14$ at the earliest emission epoch used here. The previous $1.18\times10^{-23}\,\mathrm{eV}$ benchmark would place part of the lower band below threshold and is therefore outside this example.

Differences in source amplitude or tilt between bands can mimic a change
in their auto-power ratio. The cross-spectrum tests whether the angular
patterns in the two bands align. For a common source field, replacing
the squared transfer function in Eq.~\eqref{eq:generalized_Cl_window_projection}
with the product of the two band transfers gives
\begin{equation}
C_\ell^{LH}
=4\pi\int \mathrm{d}\ln k\,
\mathcal P_{\delta S}(k)\Delta_\ell^L(k)\Delta_\ell^H(k).
\label{eq:crosscoefficient}
\end{equation}
The normalized correlation coefficient is
$r_\ell=C_\ell^{LH}/\sqrt{C_\ell^{LL}C_\ell^{HH}}$.
For massless propagation from a single shell, the two bands sample the
same angular pattern. Positive, scale-independent band amplitudes cancel
in the normalization, giving $r_\ell=1$. Different radial windows can
lower $r_\ell$ even when both bands trace the same source field.

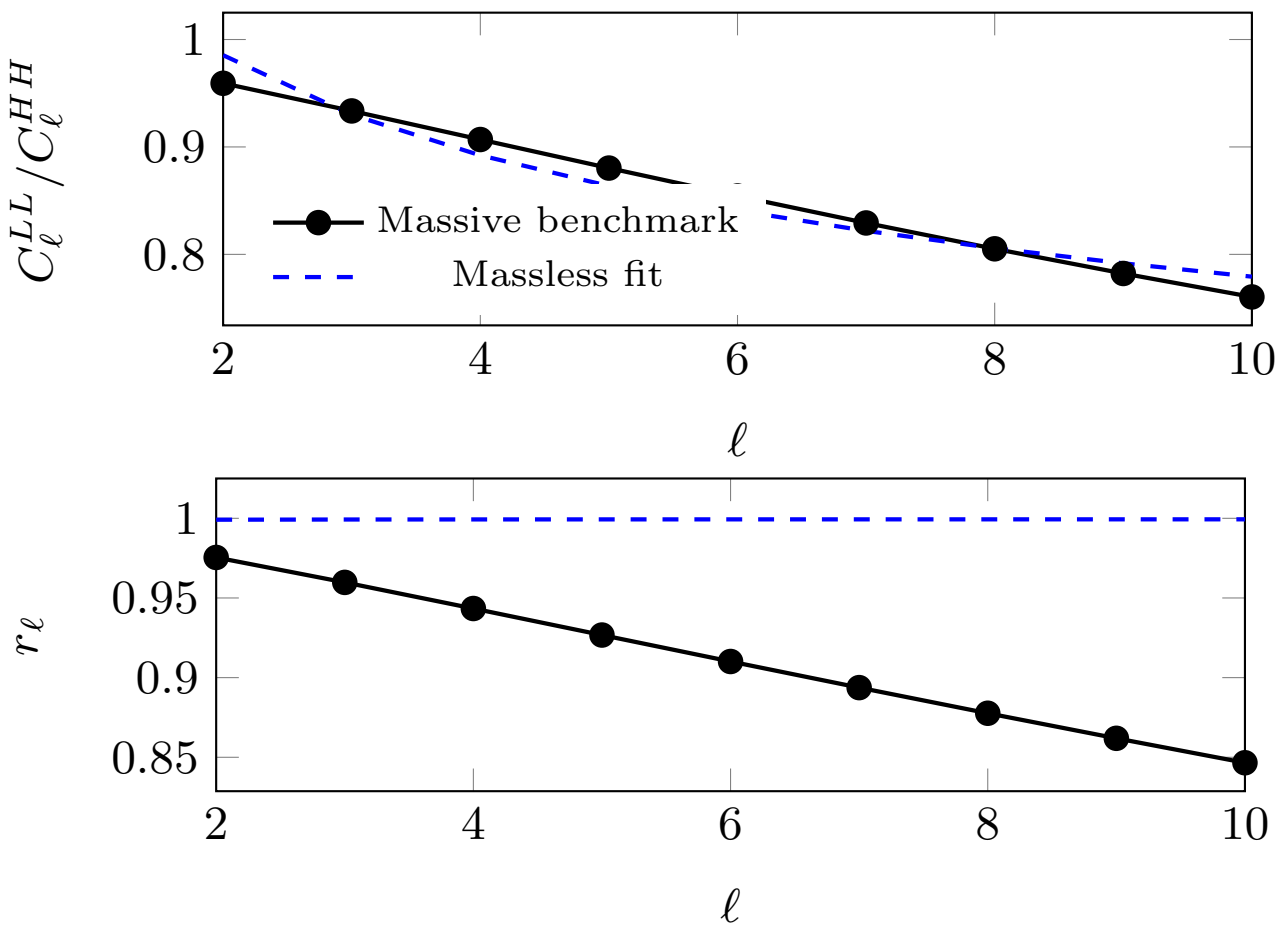
\begin{figure}[t]
\centering
\begin{tikzpicture}\begin{axis}[width=0.97\linewidth,height=3.65cm,xmin=2,xmax=10,ymax=1.025,xtick={2,4,6,8,10},xlabel={$\ell$},ylabel={$C_\ell^{LL}/C_\ell^{HH}$},legend style={font=\scriptsize,draw=none},legend pos=south west]
\addplot[black,thick,mark=*] coordinates {(2,0.9592557) (3,0.9335195) (4,0.9069031) (5,0.8803342) (6,0.8543452) (7,0.8292367) (8,0.8051670) (9,0.7822068) (10,0.7603717)};
\addplot[blue,dashed,thick] coordinates {(2,0.9854085) (3,0.9299560) (4,0.8921830) (5,0.8636620) (6,0.8408439) (7,0.8218909) (8,0.8057267) (9,0.7916664) (10,0.7792479)};
\legend{Massive benchmark,Massless fit}
\end{axis}\end{tikzpicture}\par
\begin{tikzpicture}\begin{axis}[width=0.97\linewidth,height=3.65cm,xmin=2,xmax=10,ymax=1.025,xtick={2,4,6,8,10},xlabel={$\ell$},ylabel={$r_\ell$},legend style={font=\scriptsize,draw=none},legend pos=south west]
\addplot[black,thick,mark=*] coordinates {(2,0.9754341) (3,0.9597811) (4,0.9433786) (5,0.9267291) (6,0.9101225) (7,0.8937271) (8,0.8776388) (9,0.8619094) (10,0.8465635)};
\addplot[blue,dashed,thick] coordinates {(2,0.9991181) (3,0.9992493) (4,0.9992997) (5,0.9993245) (6,0.9993385) (7,0.9993472) (8,0.9993530) (9,0.9993570) (10,0.9993600)};
\end{axis}\end{tikzpicture}\par
\caption{Two fixed bands for $m_{\rm g}=8.0\times10^{-24}\,\mathrm{eV}$ (solid) and the massless source model fitted to their auto-power ratio (dashed). The lower panel uses the same fit parameters.}
\label{fig:crossfit}
\end{figure}

For the massless comparison, we take a scale-invariant underlying field and filter it in each band with an amplitude $b_B(k/k_*)^{n_B/2}$, $b_B>0$. We fit $b_L/b_H$ and independent $n_L,n_H\in[-0.2,0.2]$ to each scale-invariant massive benchmark by minimizing the equally weighted squared residuals in $\ln(C_\ell^{LL}/C_\ell^{HH})$ over $2\leq\ell\leq10$. Define $I_\ell(n)=\int_0^\infty\mathrm dx\,x^{n-1}j_\ell^2(x)$. The cross-spectrum in Eq.~\eqref{eq:crosscoefficient} contains the product of the two filters. Setting $x=k\chi_0$ and normalizing by the auto-spectra gives
\begin{equation}
 r_\ell(0)=\frac{I_\ell[(n_L+n_H)/2]}{\sqrt{I_\ell(n_L)I_\ell(n_H)}} .
 \label{eq:tilted_null_cross}
\end{equation}
The maximum relative deviations of the fitted auto-power ratios are about $0.08\%$, $0.7\%$, and $2.7\%$, in order of increasing nonzero mass. For the heaviest case, the same massless fit predicts $r_{10}\simeq0.999$, while massive propagation gives $r_{10}\simeq0.847$ (Fig.~\ref{fig:crossfit}). This comparison uses the cross-correlation as a separate prediction of the auto-power fit.

The preceding fit assumes that both bands trace a single source field. If two independent sources contribute in different proportions to the two bands, source mixing alone can reduce their correlation even for massless propagation. Primordial modulations of a phase-transition background motivate allowing distinct early sources to have different spatial patterns~\cite{Geller2018}. To isolate this uncertainty we specify an effective emissivity model with two short episodes at $a_1=10^{-10}$ and $a_2=2\times10^{-10}$. Their fractional emissivities are modulated by independent homogeneous fields $S_1,S_2$. Their shares of the received band energy are $(w_{L1},w_{L2})=(0.8,0.2)$ and $(w_{H1},w_{H2})=(0.5,0.5)$, uniform per $\ln f$ within each band for each episode. These fractions are illustrative source inputs. A calculation of them from a specific transition would require its emission spectrum. With $\mathcal P_i(k)=A_i(k/k_*)^{n_i}$ and $\Delta_{\ell i}^B$ given by Eq.~\eqref{eq:fixed_freq_transfer} at $a_i$, independence removes the cross terms between episodes and gives
\begin{equation}
 C_\ell^{BB'}=4\pi\sum_{i=1}^2w_{Bi}w_{B'i}\int\mathrm d\ln k\,\mathcal P_i(k)\Delta_{\ell i}^B(k)\Delta_{\ell i}^{B'}(k).
 \label{eq:two_episode_power}
\end{equation}
For equal scale-invariant source powers and zero mass, the thin-shell identity makes each episode's angular power independent of radius. Equation~\eqref{eq:two_episode_power} then gives $r_\ell(0)=(\boldsymbol w_L\cdot\boldsymbol w_H)/(|\boldsymbol w_L||\boldsymbol w_H|)\simeq0.85749$. Source mixing alone therefore lowers the correlation. To compare its angular dependence we use $T=r_{10}/r_2$.

\begin{table}[t]
\centering
\small
\setlength{\tabcolsep}{7pt}
\renewcommand{\arraystretch}{1.3}
\begin{tabular}{@{}cccc@{}}
\toprule
\addlinespace[3pt]
\shortstack{$m_{\rm g}[\mathrm{eV}]$}
& $r_2$
& $r_{10}$
& $T$ \\
\addlinespace[4pt]
\midrule
0.0                  & 0.857 & 0.857 & 1.000 \\
$3.6\times 10^{-24}$ & 0.857 & 0.851 & 0.993 \\
$6.0\times 10^{-24}$ & 0.853 & 0.812 & 0.952 \\
$8.0\times 10^{-24}$ & 0.836 & 0.726 & 0.868 \\
\bottomrule
\end{tabular}
\caption{Cross-band correlations for two independent emission episodes
with equal scale-invariant source spectra. The ratio $T=r_{10}/r_2$
is evaluated before rounding.}
\label{tab:two_episode_cross}
\end{table}
For a more flexible massless comparison we keep the band fractions fixed and vary both source tilts and their relative amplitude. Write $t_\ell$ for the ratio of the two unweighted massless thin-shell powers. Then Eq.~\eqref{eq:two_episode_power} gives $r_\ell=(0.4t_\ell+0.1)/\sqrt{(0.64t_\ell+0.04)(0.25t_\ell+0.25)}$, with $t_\ell=e^g I_\ell(n_1)/I_\ell(n_2)$. The constant $g$ absorbs the source amplitudes, pivot, and shell radii. Numerical minimization and maximization over $-15\leq g\leq15$ and $-0.2\leq n_i\leq0.2$ give $0.9536\lesssim T(0)\lesssim1.0487$. The lightest benchmark in Table~\ref{tab:two_episode_cross} lies within this range. The intermediate case lies only about $0.002$ below the lower edge, whereas the heaviest lies about $0.086$ below it. Doubling the frequency quadrature from 48 to 96 nodes and the propagation quadrature from 96 to 192 nodes changed the tabulated correlations by less than $5\times10^{-6}$. The separation for the heaviest benchmark tests this specified source family near the lower-band propagation threshold. Even different constant band fractions can reproduce the same $T$. For example, taking $\boldsymbol w_L=(0.9,0.1)$, $\boldsymbol w_H=(0.1,0.9)$, $(n_1,n_2)=(-0.2,0.2)$ and $t_2=4.1122$ in the massless model gives $(r_2,r_{10})\simeq(0.273,0.237)$ and $T\simeq0.868$. The individual correlations distinguish this example from the heaviest benchmark in Table~\ref{tab:two_episode_cross}, even though their ratios agree.

A joint test follows directly from Eq.~\eqref{eq:two_episode_power}. For massless propagation each episode samples the same shell in both bands. Writing its angular power as $S_{i\ell}$ gives $C_\ell^{BB'}=\sum_i w_{Bi}w_{B'i}S_{i\ell}$. Define $X_\ell=C_\ell^{LL}/C_\ell^{HH}$ and $Y_\ell=C_\ell^{LH}/C_\ell^{HH}=r_\ell\sqrt{X_\ell}$ for $C_\ell^{HH}>0$. The vector $(X_\ell,Y_\ell,1)$ is then a linear combination of the same two vectors $(w_{Li}^2,w_{Li}w_{Hi},w_{Hi}^2)$ at every multipole. Any three such vectors are linearly dependent, giving
\begin{equation}
\det\!\begin{pmatrix}
X_i & Y_i & 1\\
X_j & Y_j & 1\\
X_k & Y_k & 1
\end{pmatrix}=0 .
\label{eq:two_source_null}
\end{equation}
Here $i,j,k$ denote multipoles. Thus the points $(X_\ell,Y_\ell)$ lie on one straight line for any two independent source powers and any band fractions fixed across multipoles. Additional independent fields, correlations between the fields, or scale-dependent mixing can invalidate the test.

For $(i,j,k)=(2,5,10)$, the heaviest massive benchmark gives a determinant of $8.96\times10^{-5}$. Increasing the frequency and propagation quadratures from 96 and 192 nodes to 192 and 384 nodes changes this value by $2.2\times10^{-8}$ or less. The nonzero determinant excludes exact agreement with the massless two-source model, even when both source powers vary freely with multipole.

To quantify how closely the massless power-law models approach the massive spectra, we fit $X_\ell$ and $r_\ell$ jointly at all nine multipoles from 2 to 10. We allow $w_{L1}$ and $w_{H1}$ to vary independently from zero to one, with $w_{B2}=1-w_{B1}$, and use the ranges for $g,n_1,n_2$ given above. Each fit minimizes the mean squared logarithm of the model-to-benchmark ratio over all 18 entries with equal weight. Differential evolution followed by multistart least squares gives root-mean-square log residuals of $0.0012$, $0.0070$, and $0.0127$ at the three increasing nonzero masses in Table~\ref{tab:two_episode_cross}. These numerical fits quantify the size of the mismatch for power-law sources. Assessing its detectability requires the covariance of the measured spectra.

\section{Conclusion}
\label{sec:conclusion}

A graviton mass changes the source radii sampled by a relic gravitational-wave background. At a fixed emission epoch, lower-momentum waves reach the observer from smaller radii, so a finite momentum band samples a shell of finite thickness. The propagation law bounds its logarithmic width by $\sigma_{\ln\chi,B}\leq\sigma_{\ln p,B}$. In a narrow, relativistic band, the fractional thickness scales as $(m_{\rm g}/p_{{\rm eff},B})^2\sigma_{\ln p,B}$. A spread in emission times also broadens the shell, including for massless propagation.

The angular-power ratio $C_\ell^{BB}(m_{\rm g})/C_\ell^{BB}(0)$ depends on the full distribution of source radii. When the momentum channels trace a common field with a scale-invariant spatial spectrum, the angular power of a thin shell is independent of its radius and bounds the power of any normalized, nonnegative radial average from above. For emission at a single epoch, the massless shell is thin, so any suppression can be attributed to radial averaging. For an extended emission history, the massless reference can itself be broad. The thin-shell bound then leaves open which mass hypothesis gives the larger angular power.

Two fixed frequency bands provide a way to test a source interpretation of the angular spectra. In the single-episode example, varying the band amplitudes and tilts in a massless model closely reproduces the massive auto-power ratio. At the largest mass, however, the same fit predicts nearly perfect correlation at $\ell=10$, whereas the massive windows sample different radii and give a lower correlation. With two independent episodes, even a massless mixture can reproduce the massive value of $r_{10}/r_2$ by changing the band fractions.

The full spectra retain a constraint that this single ratio loses. For two independent episodes with mixing fractions fixed across multipoles, the normalized auto- and cross-powers must be collinear under massless propagation, whatever the angular spectrum of each episode. The heaviest benchmark violates this condition, ruling out an exact match within that source family. The joint fits also show how small the remaining difference can be for power-law sources. Interpreting a departure from collinearity requires checking for additional source components, correlations between episodes, and scale-dependent mixing. Testing the departure with pulsar timing data further requires the mass-dependent timing response and the joint measurement covariance.

\section*{Data availability}
No observational datasets were used in this study. The curves in Figs.~\ref{fig:broadening_plot}--\ref{fig:crossfit} were calculated using the equations and parameter values given in the article. Section~\ref{sec:cross_band} describes the fit used in Fig.~\ref{fig:crossfit}. Tables~\ref{tab:frequency_mass_benchmarks}--\ref{tab:two_episode_cross} report the numerical benchmarks.

\section*{Acknowledgements}  
This work is supported by the Talent Fund of the Weihai Research Institute, Beijing Jiaotong University (2025WHRC001).

\section*{Author contributions}
H.Z. proposed the research question and wrote the manuscript.
Z.Q., S.W., J.W. and J.Z. contributed to the calculations and the preparation of the figures.

\section*{Competing interests}
The authors declare no competing interests.

\end{document}